\documentclass[11pt]{article}
\usepackage{amsmath,amssymb,color,cite}
\usepackage{graphicx}
\usepackage{amsfonts}
\usepackage{enumerate}
\usepackage{hyperref}
\usepackage{bbm}
\usepackage{nicefrac}
\usepackage[all]{xy}
\usepackage{graphicx}
\usepackage{booktabs}
\usepackage{epstopdf}
\usepackage{amsfonts}

\newcommand{\hoch}[1]{$\, ^{#1}$}
\newcommand{\cN}{{\cal N}}

\makeatletter
\@addtoreset{equation}{section}
\makeatother

\newcommand{\be}{\begin{equation}}
\newcommand{\ee}{\end{equation}}
\newcommand{\bea}{\setlength\arraycolsep{2pt} \begin{eqnarray}}
\newcommand{\eea}{\end{eqnarray}}
\newcommand{\nn}{\nonumber}

\def\ft#1#2{{\textstyle{\frac{\scriptstyle #1}{\scriptstyle #2} } }}

\def\0{{\sst{(0)}}}
\def\1{{\sst{(1)}}}
\def\2{{\sst{(2)}}}
\def\3{{\sst{(3)}}}
\def\4{{\sst{(4)}}}
\def\5{{\sst{(5)}}}
\def\6{{\sst{(6)}}}
\def\7{{\sst{(7)}}}
\def\8{{\sst{(8)}}}
\def\sst#1{{\scriptscriptstyle #1}}

\def\im{{{\rm i}}}

\begin{document}
\begin{flushright}
\hfill{ \
MI-TH-1531\ \ \ \ }
\end{flushright}

\vspace{25pt}
\begin{center}
{\Large {\bf An ${\cal N}=3$ Solution in Dyonic ISO(7) Gauged Maximal Supergravity and Its Uplift to Massive Type IIA}
}

\vspace{30pt}

{\Large
Yi Pang\hoch{1} and Junchen Rong\hoch{1}
}

\vspace{10pt}

\hoch{1} {\it George P. \& Cynthia Woods Mitchell  Institute
for Fundamental Physics and Astronomy,\\
Texas A\&M University, College Station, TX 77843, USA}

\vspace{10pt}

\vspace{20pt}

\underline{ABSTRACT}
\end{center}
\vspace{15pt}
We consider a certain $\cN=1$ supersymmetric, SO(3)$\times$SO(3) invariant, subsector of the dyonic ISO(7)-gauged maximal supergravity in four dimensions. The theory contains two scalar fields and two pseudoscalar fields. We look for stationary points of the scalar potential, especially the one preserving $\cN=3$ supersymmetry of the original ISO(7) gauged theory. The $\cN=3$ stationary point corresponding to the $AdS$ vacuum in the $D=4$ theory is lifted to a warped $AdS_4\times X_6$ type solution in massive type IIA supergravity. This $D=10$ background should be the dual of a certain $\cN=3$ Chern-Simons matter theory in three dimensions.

\thispagestyle{empty}

\pagebreak
\voffset=-40pt
\setcounter{page}{1}




\section{Introduction}
The four-dimensional ${\rm SO}(8)$ gauged maximally supersymmetric $\mathcal{N}=8$ supergravity was widely considered to be a unique theory, since its construction  thirty years ago \cite{deWit:1982ig}. Interestingly, using embedding tensor formulation \cite{deWit:2007mt}, it was recently discovered that the ${\rm SO}(8)$ gauging can be realized in a mixed electric/magnetic frame compatible with the $\cN=8$ supersymmetry \cite{Dall'Agata:2012bb}. A new parameter commonly called $\omega$ was introduced to parametrize the choice of the electric/magnetic frame. Inequivalent $\cN = 8$ theories are defined by values of $\omega$ in the interval $0\leq\omega\leq \pi/8$ \cite{Dall'Agata:2012bb, deWit:2013ija,Dall'Agata:2014ita}. This development has stimulated numerous studies, including its critical points \cite{Dall'Agata:2012bb,DallAgata:2011aa,Borghese:2012qm,Borghese:2012zs,Borghese:2013dja,Anabalon:2013eaa,lupapo,Gallerati:2014xra,Pang:2015mra} and the consequences of the deformation for the holographic dual theory \cite{Tarrio:2013qga,Guarino:2013gsa,Cremonini:2014gia,bopapose,Pang:2015mra}.
Through the work of \cite{deniwa,deWit:1986iy,deWit:2013ija,gogokrni}, it is proven that the purely electric SO(8) gauged theory can be obtained from M-theory via a consistent reduction on $S^7$. However, a no-go theorem \cite{Lee:2015xga} shows that the $\omega$-deformed SO(8) gauged supergravities cannot
be realized via a compactification that is locally described by ten- or eleven-dimensional supergravity. This means we are not able to determine the dual CFTs by applying the standard AdS/CFT correspondence, whose formulation is usually facilitated by the existence of a weakly coupled local supergravity theory on the bulk side.\footnote{It is possible that the $\omega$-deformed SO(8) gauged theories can still be embedded into string theory via a framework more general than supergravity, for example, double field theory with the strong section constraint being relaxed, following the idea of \cite{Hassler:2014sba, Blumenhagen:2014gva, Blumenhagen:2015zma}.}

Besides SO(8) gauged theory, there exist also 4D maximal supergravities with other subgroups of SL(8) being gauged. In this paper, we will consider the ISO(7) gauged theory.\footnote{Recent work studying maximal supergravities with other dyonic noncompact gauging can be seen in \cite{Kodama:2015iua, Guarino:2015tja}.} The ISO(7) gauging can be derived from the SO(8) gauging via an In\"{o}n\"{u}-Wigner contraction \cite{Hull:1984yy}. Similar to the SO(8) case, the ISO(7) group can be dyonically gauged, giving rise to two inequivalent theories: the theory in the purely electric frame and the theory in a mixed electric/magnetic frame \cite{Dall'Agata:2014ita}. The former is known to be derivable from M-theory by a consistent reduction on $S^6\times T^1$, or from type IIA supergravity via a consistent $S^6$ reduction \cite{Hull:1988jw}, the latter can be embedded to massive type IIA, with the Romans mass $m$ being identified as the parameter of the magnetic gauging upon reduction \cite{Guarino:2015jca}.
By virtue of the consistency of the reduction, solutions of the ISO(7) gauged theory can be lifted to solutions in (massive) type IIA. Some previous work on supersymmetric $AdS_4\times X_6$ type solutions in (massive) type IIA supergravity can be found in  \cite{Nunez:2001pt,Behrndt:2004km,Behrndt:2004mj,Lust:2004ig,Tomasiello:2007eq,Lust:2009mb,Rota:2015aoa,Guarino:2015jca,Fluder:2015eoa}. All these solutions preserve at most eight supercharges, or in other words, $\cN=2$ supersymmetry in four dimensions.

In this paper we present explicitly, the first analytic $AdS_4\times X_6$ solution in massive type IIA preserving $\cN=3$ supersymmetry in four dimensions. The $D=10$ solution is obtained by lifting the $\cN=3$ $AdS_4$ solution of the $D=4$ dyonic ISO(7) gauged supergravity. It was shown in \cite{Gallerati:2014xra} that the critical points in the ISO(7) gauged theory can have at most $\cN=3$ supersymmetry, and in addition to the ${\rm SO}(3)$ R-symmetry the $\cN=3$ point preserves an extra SO(3) symmetry. We shall denote the full symmetry group by ${\rm SO}(3)_R\times {\rm SO}(3)_D$.\footnote{The notation ${\rm SO}(3)_R$ should not be confused with the SO(3) R-symmetry, which is in fact associated with ${\rm SO}(3)_D$.} This subgroup is characterized by starting from
${\rm SO}(3)_1 \times{\rm SO}(3)_2\times {\rm SO}(3)_3 \times {\rm SO}(3)_4\in {\rm SO}(8)$. The factor ${\rm SO}(3)_D$ is then the diagonal in
${\rm SO}(3)_1 \times{\rm SO}(3)_2\times {\rm SO}(3)_3$, and the factor ${\rm SO}(3)_R$ is ${\rm SO}(3)_4$.\footnote{The ${\rm SO}(3)_R\times {\rm SO}(3)_D$ subgroup used here is different from the commonly used ${\rm SO}(3)\times {\rm SO}(3)$ subgroup, in which the first ${\rm SO}(3)$ refers to the diagonal in ${\rm SO}(3)_1 \times{\rm SO}(3)_2$ and the second ${\rm SO}(3)$ is the diagonal in ${\rm SO}(3)_3 \times{\rm SO}(3)_4$. Stationary points of the SO(8) and SO(4,4) gauged maximal $D=4$ supergravities which preserve the commonly used ${\rm SO}(3)\times {\rm SO}(3)$ subgroup have been uplifted to $D=11$ supergravity in \cite{Godazgar:2014eza,Baron:2014bya}.} In fact, the ${\cal N}=3$ critical point can be captured by a subsector of the full theory which is invariant under the ${\rm SO}(3)_R\times {\rm SO}(3)_D$ subgroup of ISO(7) \cite{Pang:2015mra,Guarino:2015qaa}. To obtain the $\cN=3$ critical point, we truncate the ISO(7) gauged maximal supergravity to its scalar subsector invariant under ${\rm SO}(3)_R\times {\rm SO}(3)_D$. The truncated theory preserves ${\cN} = 1$ supersymmetry, and it encompasses the scalar sectors invariant under SO(7) and ${\rm G}_2$ as special cases. The $\cN=3$ critical point appears to be a nonsupersymmetric solution of the truncated theory, however preserves $\cN=3$ supersymmetry of the original $\cN=8$ theory. Besides the $\cN=3$ critical point, we also found a single $\cN=1$, ${\rm G}_2$-invariant point and three $\cN=0$ critical points, of which the ${\rm G}_2$-and ${\rm SO}(3)_R\times {\rm SO}(3)_D$-invariant points are stable, while the SO(7) invariant point is unstable. The ${\rm G}_2$ and SO(7)-invariant point is previously known in \cite{Borghese:2012qm}, where the ${\rm G}_2$ and SO(7)-invariant points in $D=4$ maximal supergravities with all gaugings are listed. We then uplift the $\cN=3$ critical point as well as the ${\rm G}_2$-invariant, $\cN=1,0$ critical points to massive type IIA supergravity using the uplift formulas given in \cite{Guarino:2015jca}.

This paper is organized as follows. In Sec. \ref{embedding}, we briefly review the ingredients of the dyonic ISO(7) gauged supergravity. In Sec. \ref{SO(3)SO(3)sector}, we truncate the ISO(7) gauged supergravity to the scalar subsector invariant under the ${\rm SO}(3)_R\times {\rm SO}(3)_D$, list the critical points in this sector, and study the mass spectra of the scalar fluctuations around these $AdS$ vacua. In Sec. \ref{uplift}, we uplift the $D=4$ critical points to $D=10$ and obtain the first $\cN=3$ solution in massive type IIA explicitly. We discuss possible CFT dual of the $\cN=3$ solution and conclude in Sec. \ref{CFTdual}.

{\it Note Added}: While we were preparing this draft, the work \cite{Guarino:2015qaa} appeared
in arXiv and it seems to overlap with our paper on the critical points and their spectra analysis in the ${\rm SO}(3)_D\times {\rm SO}(3)_R$-invariant sector.

\section{Dyonic ISO(7) gauged $\cN=8$ supergravity}\label{embedding}
The $ D=4$ gauged maximal supergravity is characterized by the embedding tensor $\Theta_M{}^{\alpha}$ which completely specifies the gauging \cite{deWit:2007mt}. The embedding tensor enters the Lagrangian through the quantity
\be
X_M=\Theta_M{}^{\alpha}t_{\alpha},
\ee
which satisfies
\be
[X_M,X_N]=-X_{MN}{}^P X_P,
\ee
implying a closed algebra. Indices $M, N\ldots$ transform as the $\bf{56}$ of ${\rm E}_{7(7)}$ which decomposes into ${\bf 28 }\oplus \overline{{\bf 28 }}$ of ${\rm SU}(8)$ or ${\bf 28 }\oplus {\bf 28 }'$ of SL(8). The decomposition of the $\bf{56}$ of ${\rm E}_{7(7)}$ under ${\rm SU}(8)$ or SL(8) suggests two different bases in which ${\rm E}_{7(7)}$ covariant quantities can be formulated. In the SL(8) basis, $V_M=\{V_{[AB]},V^{[AB]}\}$, and the pure scalar sector of the 4D gauged maximal supergravity is given as
\be
e^{-1}{\cal L}=\frac{1}{8}\text{Tr}(\partial_{\mu}\mathcal{M}\partial^{\mu}\mathcal{M}^{-1})
-\frac{1}{672}(X_{MN}{}^RX_{PQ}{}^S\mathcal{M}^{MP}\mathcal{M}^{NQ}\mathcal{M}_{RS}+7X_{MN}{}^Q X_{PQ}{}^N \mathcal{M}^{MP}),
\label{la1}
\ee
where $\mathcal{M}^{MN}$ is the inverse of $\mathcal{M}_{MN}$. The latter is constructed from the bilinear of the 56-bein
\be
\mathcal{M}_{MN}=(LL^{\dagger})_{MN},\quad L(\phi)_{M}{}^{\underline{N}}=S^{\dagger}_M{}^{\underline{P}}\hat{L}(\phi)_{\underline{P}}{}^{{\underline{N}}},
\label{scalarmatrix}
\ee
with
\be
S_{\underline{M}}{}^{N}=\frac1{4\sqrt{2}}
\left(\begin{array}{cc}
\Gamma_{ij}{}^{AB} & \im\Gamma_{ijAB} \\
\Gamma^{ijAB} & -\im\Gamma^{ij}{}_{AB}
\end{array}\right),\quad \hat{L}_{\underline{M}}{}^{\underline{N}}=\exp\left(\begin{array}{cc}
0 & \phi_{ijkl} \\
\phi^{ijkl} & 0
\end{array} \right).
\label{cosetrep}
\ee
The indices $\underline{M},\underline{N}$ label the ${\bf 56}$ irreps realized in the ${\rm SU}(8)$ basis, and therefore the unitary matrix $S^{\dagger}$ converts the ${\rm SU}(8)$ basis to the SL(8) basis \cite{DallAgata:2011aa}. Indices $A, B,\ldots$ are associated with the fundamental representation of SL(8) while $i,j,\ldots$ transform as the ${\bf 8}$ of ${\rm SU}(8)$. The 2-gamma matrices $\Gamma_{ij}{}^{AB}$ comprise the generators of ${\rm SO}(8)$ in the chiral spinor representation. The position of the index is not crucial. Here, we use the same notation as \cite{Pang:2015mra} where the convention of the 2-gamma matrices is given explicitly.\footnote{Slightly different from \cite{Pang:2015mra}, here we interchanged the definition of $\Gamma_2$ and $\Gamma_7$.}

Dyonic ISO(7) gauged maximal supergravity arises when the electric and magnetic parts of the embedding tensor are both nonvanishing. Specifically,
$X_{MN}{}^P$ takes the form
\be
X_{ABM}{}^N=\left(\begin{array}{cc}
-f_{ABCD}{}^{EF} &  \\
 & f_{ABEF}{}^{CD}
\end{array}
\right),\quad
X^{AB}{}_{M}{}^N=\left(\begin{array}{cc}
-g^{AB}{}_{CD}{}^{EF} &  \\
 & g^{AB}{}_{EF}{}^{CD}
\end{array}
\right),
\ee
where $f_{ABCD}{}^{EF}=2\sqrt{2}\delta^{[E}_{[A}\theta_{B][C}\delta_{D]}^{F]}$ and $g^{AB}{}_{EF}{}^{CD}=2\sqrt{2}\delta_{[E}^{[A}\xi^{B][C}\delta^{D]}_{F]}$ in which
 \be
\theta=g\cdot\text{diag}(\mathbb{I}_7,0),\quad \xi=m\cdot\text{diag}(0_7,1).
\ee
The inequivalent theories correspond to $m=0$ or $m\neq0$. The reason \cite{Dall'Agata:2014ita} is that one can use the SL(8) Cartan generator,
\be
\Lambda_{\rm red}=\left(\begin{array}{cc}
\mathbb{I}_7 &  \\
 & -7
\end{array}
\right),
\label{4dscaling}
\ee
to rescale the electric and magnetic coupling constant $(g,m)$ separately, and the effects of $\Lambda_{\rm red}$ on the scalar matrix $\mathcal{M}_{MN}$ can be absorbed by a nonlinear field redefinition.

\section{Truncation to the ${\rm SO}(3)_R\times {\rm SO}(3)_D$ invariant sector}\label{SO(3)SO(3)sector}
In this section, we truncate the ${\rm ISO}(7)={\rm SO}(7)\times \mathbb{R}^7$ gauged theory to its subsector invariant under an ${\rm SO}(3)_R\times {\rm SO}(3)_D$ group, which is embedded in ISO(7) by the following series:
\be
{\rm ISO}(7)\supset {\rm SO}(7)\supset {\rm SO}(3)_R\times {\rm SO}(3)_L \times {\rm SO}(3)\supset {\rm SO}(3)_R\times \big[{\rm SO}(3)_L \times {\rm SO}(3)\big]_D
\ee
Here $\big[{\rm SO}(3)_L \times {\rm SO}(3)\big]_D$ means the diagonal subgroup of ${\rm SO}(3)_L \times {\rm SO}(3)$, which we denote by ${\rm SO}(3)_D$ in later discussion for brevity. The scalar coset invariant under the ${\rm SO}(3)_R\times {\rm SO}(3)_D$ subgroup can be parametrized by the following ${\rm SU}(8)$ complex self-dual 4-form:\footnote{One way of understanding this formula is to treat ${\rm SO}(3)_R\times {\rm SO}(3)_D$ as a subgroup of SO(8). In the of notation the gamma matrices, $\{i, A, \dot{A}\}$ label the indices of $\{8_{\rm s}, 8_{\rm v}, 8_{\rm c}\}$ representations respectively. $\Phi_{1AB}$ and $\Phi_{2AB}$  break ${\rm SO}(8)$ into $[{\rm SO}(3)\times {\rm SO}(4)]_{\rm v}$, while $\Phi_{1\dot{A}\dot{B}}$ and $\Phi_{2\dot{A}\dot{B}}$ break ${\rm SO}(8)$ into $[{\rm SO}(3)\times {\rm SO}(4)]_{\rm c}$. When $\Phi_{1AB},\, \Phi_{2AB},\, \Phi_{1\dot{A}\dot{B}},\, \Phi_{2\dot{A}\dot{B}}$ are present at the same time, ${\rm SO}(8)$ is broken into the ${\rm SO}(3)_R\times {\rm SO}(3)_D$ subgroup, which is the common subgroup of $[{\rm SO}(3)\times {\rm SO}(4)]_{\rm v}$ and $[{\rm SO}(3)\times {\rm SO}(4)]_{\rm c}$.}
\bea
\phi_{ijkl}&=&\left(\phi_1 \cos\sigma_1 \Phi_1+ \phi_2 \cos\sigma_2\Phi_2\right)_{AB} (\Gamma^{ijkl})_{AB}\nn\\
&&+\im(\phi_1 \sin\sigma_1\Phi_1+ \phi_2 \sin\sigma_2\Phi_2)_{\dot{A}\dot{B}} (\Gamma^{ijkl})_{\dot{A}\dot{B}},
\eea
where
\be
\Phi_1=\frac{1}{2}\text{diag}(-1,-1,-1,-1,1,1,1,1),\quad \Phi_2=\text{diag}(0,0,0,0,-1,-1,-1,3),
\ee
and $(\Gamma^{ijkl})_{AB}$, $(\Gamma^{ijkl})_{\dot{A}\dot{B}}$ are the upper and lower 8 by 8 diagonal blocks of the 16 by 16 4-gamma matrix respectively
\be
\Gamma^{ijkl}=\left(\begin{array}{cc}
(\Gamma^{ijkl})_{AB} &  \\
 & (\Gamma^{ijkl})_{\dot{A}\dot{B}}
\end{array}\right).
\ee
Substituting the scalar 4-form ansatz into the Lagrangian \eqref{la1}, we obtain the scalar potential for the ${\rm SO}(3)_D\times {\rm SO}(3)_R$ invariant sector of the dyonic ISO(7) gauged maximal supergravity as
\begin{eqnarray}
V=&&\frac{1}{64} g^2 (\cosh\phi _2-\cos\sigma _2 \sinh\phi _2)^2 (\cos\sigma _2 \sinh\phi _2+\cosh\phi _2)\nonumber\\&&
\times  \Big(196 \cos(\sigma _1+3 \sigma _2) \sinh\phi _1\sinh ^3\phi _2+4 \cos(\sigma _1-3 \sigma _2) \sinh\phi _1 \sinh ^3\phi _2\nonumber\\&&
+35 \cos(\sigma _1-\sigma _2) \sinh\phi _1 \sinh\phi _2+467 \cos(\sigma _1+\sigma _2) \sinh\phi _1 \sinh\phi _2\nonumber\\
&&+47 \cos(\sigma _1-\sigma _2) \sinh\phi _1 \sinh 3\phi _2-97 \cos(\sigma _1+\sigma _2) \sinh \phi _1 \sinh 3\phi _2\nonumber\\
&&+4 \cos \sigma _1 \sinh \phi _1 \cosh \phi _2(28 \cos 2\sigma _2 \sinh ^2\phi _2+47)\nonumber\\
&&-2 \cosh \phi _1 (28 \cos 3\sigma _2 \sinh ^3\phi _2+101 \cos \sigma _2 \sinh \phi _2-7 \cos \sigma _2 \sinh 3\phi _2\nonumber\\
&&+ \cosh \phi _2(200 \cos 2\sigma _2 \sinh ^2\phi _2+226)-50 \cosh 3 \phi _2)\nonumber\\
&&-28 \cos \sigma _1 \sinh \phi _1 \cosh 3 \phi _2-768\Big)\nonumber\\
&&+g m \sin ^2\sigma _2 \sinh ^2\phi _2 (\cos \sigma _2 \sinh \phi _2-\cosh\phi _2)^3\times\Big(3 \sin \sigma _1 \sinh \phi _1 \cosh \phi _2\nonumber\\
&&+\sinh \phi _2 \left(4 \sin \sigma _2 \cosh \phi _1-\sinh \phi _1 \left(3 \sin \sigma _1 \cos\sigma _2+4 \sin \sigma _2 \cos \sigma _1\right)\right)\Big)\nonumber\\
&&+\frac{1}{2} m^2 \left(\cosh \phi _1-\cos \sigma _1 \sinh \phi _1\right) \left(\cosh \phi _2-\cos \sigma _2 \sinh \phi _2\right)^6.
\label{sp}
\end{eqnarray}
So far we have been focused only on the scalar sector. In fact, the full fledged ${\rm SO}(3)_D\times {\rm SO}(3)_R$ invariant sector preserves $\cN=1$ supersymmetry which implies the potential can be expressed in terms of a superpotential $W$, with
\be
V=2\Big(4|\frac{\partial W}{\partial \phi_1}|^2+\frac23|\frac{\partial W}{\partial \phi_2}|^2-3|W|^2\Big),
\ee
where
\bea
W&=&\frac{g}4(\cosh\frac{\phi_2}2+e^{\im \sigma_2}\sinh\frac{\phi_2}2)^2(\cosh\frac{\phi_2}2-e^{\im \sigma_2}\sinh\frac{\phi_2}2)^3\nn\\
&&\times \Big(\cosh\frac{\phi_2}2(7\cosh\frac{\phi_1}2-e^{\im \sigma_1}\sinh\frac{\phi_1}2)+e^{\im\sigma_2}(\cosh\frac{\phi_1}2-7e^{\im \sigma_1}\sinh\frac{\phi_1}2)\sinh\frac{\phi_2}2\Big)\nn\\
&&+\frac{\im m}4(\cosh\frac{\phi_1}2-e^{\im \sigma_1}\sinh\frac{\phi_1}2)(\cosh\frac{\phi_2}2-e^{\im \sigma_2}\sinh\frac{\phi_2}2)^6.
\eea
The location of the critical point depends on $m/g$, however, as mentioned before, the effects due to different $m/g$ can be compensated by a nonlinear field redefinition. Therefore we can choose this ratio to be the one most convenient for our purpose. In finding the locations of the critical points, we choose
\be
\frac{m}{g}=2.
\ee
The scalar potential \eqref{sp} possesses an ${\rm SO}(3)_R \times {\rm SO}(3)_D$ invariant stationary point preserving $\cN=3$ supersymmetry of the original $\cN=8$ theory which lies outside the residual $\cN=1$ supersymmetry of the truncated theory.
In terms of the complexified fields
\be
\xi_1=\tanh\phi_1e^{\im \sigma_1},\quad \xi_2=\tanh\phi_2e^{\im \sigma_2},
\ee
the $\cN=3$ point is given by
\be
\xi_1=\frac{3}{5}-\frac{2\im}{5},\quad \xi_2=\frac{\im}{2},
\ee
The mass spectrum of the fluctuations around this vacuum has been obtained previously in
\cite{Gallerati:2014xra} by a group theoretic method without referring to the detailed position of
the critical point:
\begin{eqnarray}
m^2 L_0^2 & : & \quad 1\times(3(1+\sqrt{3})); \quad
 6\times(1+\sqrt{3});\quad 1\times(3(1-\sqrt{3}));\quad
 6\times(1-\sqrt{3});\nonumber\\
&&\quad 4\times(-\frac{9}{4});\quad
 18\times(-2);\quad 12\times(-\frac{5}{4});\quad 22\times 0,
 \label{n=3spectrum}
\end{eqnarray}
where the $AdS$ radius squared $L^{2}_0=-6/V$, $V(g=1,m=2)=-32/\sqrt{3}$.
(The integer to the left of the multiplication sign indicates the
degeneracy of the mass eigenvalue, while the number to the right indicates
the corresponding mass squared.)
There is another $\cN=1$ critical point with ${\rm G}_2$ global symmetry:
\be
\xi_1=\xi_2=-\frac{\im}{4}.
\ee
To obtain this critical point, we have combined the Newton-Raphson method with the ``inverse Symbolic Calculator" technique. The mass spectrum of the scalar fluctuations around this vacuum is given by
\be
m^2 L_0^2 : \quad 1\times(4\pm\sqrt{6}); \quad
 14\times 0;\quad 27\times (-\ft16(11-\sqrt{6})),
\ee
where the $AdS$ radius squared $L^{2}_0=-6/V$, $V(g=1,m=2)=-512\sqrt{3}/(25\sqrt{5})$.
Besides the supersymmetric critical points,
there are three more nonsupersymmetric critical points, preserving ${\rm SO}(3)_R \times {\rm SO}(3)_D$, ${\rm G}_2$ and ${\rm SO}(7)_{\rm v}$ global symmetries respectively.
The ${\rm SO}(3)_R \times {\rm SO}(3)_D$ invariant critical point is located at
\be
\xi_1=0.353669 +0.0552267 \im,\quad \xi_2=-0.0293804 + 0.534729 \im,
\ee
with the mass spectrum given by
\begin{eqnarray}
m^2 L_0^2 & : & \quad 1\times(6.72740);\quad 1\times(5.28662);
\quad 4\times(-1.96422);\quad 9\times(-1.75110);\nonumber\\
&&\quad 9\times(-1.58816);\quad 1\times(-1.58552);\quad 8\times(-1.17591);
\quad 5\times(-0.98271);\nonumber\\
&&\quad 4\times(-0.72962);\quad 5\times(0.62977);\quad
1\times(0.58436);\quad 22\times 0,
\end{eqnarray}
where the $AdS$ radius squared $L^{2}_0=-6/V$, $V(g=1,m=2)=-18.662034$.
The ${\rm G}_2$ invariant point is given by
\be
\xi_1=\xi_2=\frac{\im}{2}.
\ee
The associated mass spectrum has the simple structure
\be
m^2 L_0^2 : \quad 2\times 6; \quad
 14\times 0;\quad 54\times (-1),
\ee
where the $AdS$ radius squared $L^{2}_0=-6/V$, $V(g=1,m=2)=-32/\sqrt{3}$.
It should be emphasized that the spectra associated with the nonsupersymmetric ${\rm SO}(3)_R \times {\rm SO}(3)_D$-and ${\rm G}_2$-invariant critical points lie above the Breitenlohner-Freedman (BF) bound. The last ${\rm SO}(7)_{\rm v}$-invariant critical point is located at
\be
\phi_1=\phi_2=-\ft16\log\ft54,\quad \sigma_1=\sigma_2=0.
\ee
The mass spectrum of the ${\rm SO}(7)_{\rm v}$-invariant critical point reads
\be
m^2 L_0^2 : \quad 1\times 6; \quad
 7\times 0;\quad 35\times (-\ft65);\quad 27\times (-\ft{12}5),
\ee
where the $AdS$ radius squared $L^{2}_0=-6/V$, $V(g=1,m=2)= -15\times5^{\frac16}/2^{\frac13}$.
This point is unstable against fluctuations as the mass squared of some scalar modes is below the BF bound. The scalar mass spectra associated with the two ${\rm G}_2$-invariant critical points and the single ${\rm SO}(7)_{\rm v}$-invariant point coincide with those given in \cite{Borghese:2012qm}, where the ${\rm G}_2$- and ${\rm SO}(7)_{\rm v}$-invariant critical points in $D=4$ maximal supergravities with all gaugings are analyzed.

Since different choices of $m/g$ are related by highly nonlinear field redefinition, we expect that if $m/g$ had been different from 2, the expressions of the critical points will not be as neat as the ones given above.
\section{Lifting the 4D critical points to 10D massive type IIA}\label{uplift}
Previously, in order to find the exact positions of the critical points, we have set $m=2g$ for convenience. Thus a direct application of the uplift formulas will lead to solutions in massive type IIA supergravity with $m=2g$. To recover the general $m$-dependence of the solution, we utilize two scaling symmetries of the equations of motion of the massive type IIA supergravity. In the type IIA Einstein frame conventions of \cite{Cvetic:1999un}, these scalings act on the fields and the parameter of the theory as follows
\bea
&&\hat{A}_{(1)}\rightarrow \tau \hat{A}_{(1)},\quad \hat{A}_{(2)}\rightarrow \tau \hat{A}_{(2)},\quad \hat{A}_{(3)}\rightarrow \tau^2 \hat{A}_{(3)},\quad d\hat{s}^2_{10}\rightarrow \tau^{\ft54}d\hat{s}^2_{10},\quad e^{\hat{\phi}}\rightarrow \tau^{-\ft12}e^{\hat{\phi}},\nn\\
&&\hat{A}_{(1)}\rightarrow \kappa \hat{A}_{(1)},\quad \hat{A}_{(2)}\rightarrow \kappa^2 \hat{A}_{(2)},\quad \hat{A}_{(3)}\rightarrow \kappa^3\hat{A}_{(3)},\quad d\hat{s}^2_{10}\rightarrow \kappa^{2} d\hat{s}^2_{10},\quad m\rightarrow \frac{m}{\kappa}.
\label{scalings}
\eea
Notice that the second scaling is merely based on the dimensionality. The scaling symmetry of the 4D theory (\ref{4dscaling}) reflects itself in the 10D theory as a combination of the above two scalings with $\tau=\lambda^{20}, \kappa=\lambda^{-14}$, where $\lambda$ is identified as the parameter of $\Lambda_{\rm red}$ (\ref{4dscaling}).

\subsection{Supersymmetric ${\rm SO}(3)_R\times {\rm SO}(3)_L$ invariant solution in massive IIA}
In terms of the auxiliary coordinates on $S^6$
\bea
\mu^1&=&\sin \xi \cos \theta _1  \cos \chi _1,\quad \mu^2=\sin \xi \cos \theta _1  \sin \chi _1,\nonumber\\
\mu^3&=&\sin \xi \sin \theta _1  \cos \psi ,\quad \mu^4= \sin \xi \sin \theta _1\sin \psi ,\nonumber\\
\nu^1&=&\cos \xi  \cos \theta _2 ,\quad \nu^2=\cos \xi  \sin \theta _2  \cos \chi _2,\nonumber\\
\nu^3&=&\cos \xi  \sin \theta _2  \sin \chi _2,
\eea
which satisfy $\sum_{A=1}^{4}\mu^A\mu^A+\sum_{i=1}^{3}\nu^{i}\nu^{i}=1$, the metric on the round $S^6$ takes the form
\bea\label{spheremetric}
ds_{S^6}^2&=& d\xi^2+\sin^2\xi d\Omega_3^2+\cos^2\xi d\Omega_2^2\nonumber\\
&=&d\xi^2+\sin^2\xi\left(d\theta_1^2+\cos^2\theta_1d\chi_1^2+\sin^2\theta_1d\psi^2\right)+\cos^2\xi(d\theta_2^2+\cos^2\theta_2d\chi_2^2).
\eea
To lift the solution of the 4D dyonic ISO(7) gauged supergravity to that in the 10D massive type IIA supergravity, we utilize the uplift formulas given in \cite{Guarino:2015jca}, in which the internal components of the 10D metric, the dilaton, and various form fields are constructed in terms of the SL(7)-covariant blocks of the $D=4$ scalar matrix ${\cal M}_{MN}$:
\bea\label{invesemetric}
g^{mn}&=&\frac{1}{4}g^2 \Delta K^{m}_{IJ}  K^{n}_{KL}\mathcal{M}^{IJ,KL},\nn\\
e^{-\frac{3}{2}\hat{\phi}}&=&-g^{mn}\hat{A}_m \hat{A}_n+\Delta x_I x_J \mathcal{M}^{I8J8},\nn\\
\hat{A}_m&=&\frac{1}{2} g \Delta g_{mn} K^{n}_{IJ} x_K \mathcal{M}^{IJK8},\nn\\
\hat{A}_{mn}&=&-\frac{1}{2}\Delta g_{pm} K^{p}_{IJ} \partial_n x^K \mathcal{M}^{IJ}{}_{K8},\nn\\
\hat{A}_{mnp}&=& \hat{A}_m \hat{A}_{np}+\frac{1}{8}g \Delta g_{mq} K^{q}_{IJ} K_{np}^{KL}\mathcal{M}^{IJ}{}_{KL},
\eea
where $K^{m}_{IJ}=2g^{-2}\mathring{g}^{mn}x_{[I}\partial_n x_{J]}$, $K_{mn}^{IJ}=4 g^{-2} \partial_{[m}x_{I}\partial_{n]}x_{J}$. Due to our gamma matrix notation, $\{x^I\}$ are related to $\{\mu^A, \nu^i\}$ by the similarity transformation
\be
{\cal S}=\text{diag}(1,1,1,1,-1,-1,1),
\ee
which brings the SL(7)-covariant blocks of the $D=4$ scalar matrix ${\cal M}_{MN}$ into a form invariant under the standard ${\rm SO}(3)_R\times {\rm SO}(3)_D$ transformation given in \cite{Pang:2015mra}.

The 10D solution corresponding to the $\cN=3$ critical point is then obtained as follows:
\be
L^{-2}d\hat{s}^2_{10}= \Delta^{-1}(\frac{3\sqrt{3}}{16}ds^2_{AdS_4})+g_{mn} dy^m dy^n,
\ee
where
\be
\Delta=3^{\ft98}2^{-\ft34}(\cos 2 \xi +3)^{-\ft18} \Xi^{-\ft14},\qquad \Xi=(24 \cos 2 \xi +3 \cos 4 \xi +37),
\ee
and the internal metric on the deformed $S^6$ is given as
\bea
g_{mn}dy^m dy^n&=\frac{3\sqrt{3}}{4}(\Delta{\Xi})^{-1} \bigg[&-\sin ^2 2 \xi  d\xi^2+8 (\cos 2 \xi +3) d\mu \cdot d\mu+4 (\cos 2 \xi +3)d\nu\cdot d\nu\nonumber\\&&+16 \mu^A \eta^i_{AB} d\mu^B\epsilon^{ijk}\nu^jd\nu^k-\frac{16}{\cos 2 \xi +3} (d\mu^A \eta^i_{AB} \mu^B \nu^i)^2\bigg],
\eea
where $\eta^i$'s are the generators of ${\rm SO}(3)_L$ embedded in ${\rm SO}(4)\simeq {\rm SO}(3)_R\times {\rm SO}(3)_L$,
\be
\eta^1=\left(
\begin{array}{cccc}
 0 & 0 & 0 & -1 \\
 0 & 0 & -1 & 0 \\
 0 & 1 & 0 & 0 \\
 1 & 0 & 0 & 0 \\
\end{array}
\right),\quad
\eta^2=\left(
\begin{array}{cccc}
 0 & 0 & 1 & 0 \\
 0 & 0 & 0 & -1 \\
 -1 & 0 & 0 & 0 \\
 0 & 1 & 0 & 0 \\
\end{array}
\right),\quad
\eta^3=\left(
\begin{array}{cccc}
 0 & -1 & 0 & 0 \\
 1 & 0 & 0 & 0 \\
 0 & 0 & 0 & -1 \\
 0 & 0 & 1 & 0
\end{array}
\right).
\ee
Denote $\mathcal{K}^{i}\equiv\mu^A \eta^i_{AB} d\mu^B$, various p-form fields can then be expressed as
\bea
L^{-1}e^{\frac{3}{4}\phi_0}\hat{{A}}_{(1)}&= &\frac{2}{\cos 2 \xi +3} \nu^i \mathcal{K}^i, \\
L^{-2}e^{-\frac{1}{2}\phi_0}\hat{{A}}_{(2)}&=&- \Xi^{-1}\bigg[-8 d\nu^i\wedge\mathcal{K}^i+ 6 \sin 2 \xi d\xi \wedge \nu^i \mathcal{K}^i+2 (3 \cos 2 \xi +5)\nu^i \eta^i_{AB}d\mu^A \wedge d\mu^B\nonumber\\
&&\qquad\qquad  -(3 \cos 2 \xi +5) \epsilon^{ijk} \nu^i d\nu^j\wedge d\nu^k \bigg],\\
L^{-3}e^{\frac{1}{4}\phi_0}\hat{{A}}_{(3)}&=&-\Xi^{-1}\bigg[6 \sin 2 \xi  \epsilon^{ijk} d\xi \wedge \mathcal{K}^i \nu^j \wedge d\nu^k+2 (3 \cos 2 \xi +5)\epsilon^{ijk} \nu^i d\nu^j\wedge d\mu^A \wedge \eta^k_{AB} d\mu^B\nonumber\\
&&\qquad+4\epsilon^{ijk} \mathcal{K}^i\wedge d\nu^j\wedge d\nu^k+\frac{8}3 \csc ^2\xi \epsilon^{ijk} \mathcal{K}^i\wedge \mathcal{K}^j \wedge \mathcal{K}^k\bigg]+\frac{ 3\sqrt{3}}{8}\Omega_{(3)}.
\eea
where $d\Omega_{(3)}={\rm vol}(AdS_4)$ which is the volume element of the ``unit" $AdS_4$.
Finally, the 10D dilaton is given by
\be
e^{-\frac{3}{2}\hat{\phi}}=e^{-\frac{3}{2}\phi_0}\frac{\Delta\Xi}{3\sqrt{3}(\cos 2 \xi +3)}.
\ee
Notice that everything is written in terms of ${\rm SO}(3)_R\times {\rm SO}(3)_D$ invariant quantities (any function of $\xi$ is invariant as $\mu\cdot\mu=\sin^2\xi$ is an invariant quantity ). Here we have introduced two constants $L^2=2^{-\frac{1}{12}}g^{-25/12}m^{1/12}$ and $e^{\phi_0}=2^{\frac{5}{6}}g^\frac{5}{6}m^{-\frac{5}{6}}$ using the scaling symmetries \eqref{scalings}. We have checked that our solution satisfies all the equations of motion of massive type IIA supergravity in the convention of \cite{Cvetic:1999un}.

\subsection{${\rm G}_2$-invariant solutions in massive type IIA }
In our notation, we can write down the almost complex structure on unit $S^6$ as
\be
J_{(2)}=\frac{1}{2!} J_{mn}dy^m\wedge dy^n= \mathcal{K}^i \wedge d\nu^i+\frac{1}{2} \nu^i \eta^i_{AB}  d\mu^A \wedge d\mu^B+\frac{1}{2}\epsilon^{ijk}\nu^i d\nu^j \wedge d\nu^k,
\ee
which satisfies $J_{mn}J^{nl}=-\delta_m^l$, and also $-\frac{1}{2}J_{(2)}\wedge J_{(2)}=*_6J_{(2)}$. The parallel torsion of $J_{(2)}$ is
\be
G_{(3)}=-\frac{1}{3} dJ_{(2)}.
\ee
Then $H_{(3)}\equiv*_6G_{(3)}$ satisfies the relation
\be
dH_{(3)}=2 J_{(2)}\wedge J_{(2)},
\ee
where ``$*_6$'' is the Hodge dual defined with respect to the $S^6$ metric. The uplift of the $\cN=1$ ${\rm G}_2$-invariant critical point gives rise to the 10D solution
\bea
L^{-2}d\hat{s}^2_{10}&=& \alpha^{-3}\big(\frac{25 \sqrt{15}}{256} ds^2_{AdS_4}\big)+\alpha ds^2_{S^6},\quad \alpha=\frac{15^{3/8}}{2 \sqrt{2}},\nn\\
e^{-\frac{3}{2}\hat{\phi}}&=&e^{-\frac{3}{2}\phi_0}\alpha^{-1},\qquad L^{-1}e^{\frac{3}{4}\phi_0} \hat{A}_{(1)}= 0,\nn\\
L^{-2}e^{-\frac{1}{2}\phi_0}\hat{A}_{(2)}&=& \frac{1}{4}J_{(2)},\qquad L^{-3}e^{\frac{1}{4}\phi_0} \hat{A}_{(3)}= \frac{1}{4} H_{(3)} +\frac{25 \sqrt{15}}{128}  \Omega_{(3)},
\eea
where $ds^2_{S^6}$ is the metric of the unit $S^6$ given in (\ref{spheremetric}). Again, we introduced $L^2=2^{-\frac{1}{12}}g^{-25/12}m^{1/12}$ and $e^{\phi_0}=2^{\frac{5}{6}}g^\frac{5}{6}m^{-\frac{5}{6}}$ using the scaling symmetries. It is recalled that in \cite{Behrndt:2004mj}, general $\cN=1$\footnote{Here $\cN=1$ means four real supercharges.} flux compactification of massive type IIA string has been analyzed. Therefore, our $\cN=1$, ${\rm G}_2$-invariant solution should be contained in the discussion of \cite{Behrndt:2004mj}.

The stable nonsupersymmetric ${\rm G}_2$-invariant solution is obtained by uplifting the nonsupersymmetric ${\rm G}_2$-invariant critical point of the $D=4$ theory and the result is given by
\bea
L^{-2}d\hat{s}^2_{10}&=&\alpha^{-3}\big(\frac{3 \sqrt{3}}{16} ds^2_{AdS_4}\big)+\alpha ds^2_{S^6},\quad \alpha=\frac{3^{3/8}}{2^{3/4}},\nn\\
e^{-\frac{3}{2}\hat{\phi}}&=&e^{-\frac{3}{2}\phi_0}\alpha^{-1},\qquad L^{-1}e^{\frac{3}{4}\phi_0}\hat{ A}_{(1)}= 0,\nn\\
L^{-2}e^{-\frac{1}{2}\phi_0} \hat{A}_{(2)}&=& -\frac{1}{2}J_{(2)}, \qquad L^{-3}e^{\frac{1}{4}\phi_0} \hat{A}_{(3)}= -\frac{1}{2} H_{(3)} + \frac{3 \sqrt{3}}{8} \Omega_{(3)}.
\eea
\section{Discussions and conclusions}\label{CFTdual}
In the paper \cite{Guarino:2015jca}, it has been proposed that the holographic duals of the $AdS_4\times \widetilde{S}^6$ ($\widetilde{S}^6$ means a deformed 6-sphere) backgrounds in massive type IIA are the simplest superconformal field theory first considered by Schwarz \cite{Schwarz:2004yj}, with a simple gauge group ${\rm SU}(N)$, adjoint matter and level $k$. The Romans mass parameter $m$ which gives rise to the magnetic coupling of the $D=4$ theory has been argued to be related to the Chern-Simons level by \cite{Gaiotto:2009mv, Guarino:2015jca}
\be
m=k/(2\pi \ell_s),
\ee
where $\ell_s$ is the string length. The ${\cN}=2$ solutions found in \cite{Guarino:2015jca,Fluder:2015eoa} are conjectured to be dual to certain ${\cN}=2$ Chern-Simons matter theories. Some evidence has been given by matching the partitions of the supergravity solutions with those of the conjectured CFT's \cite{Guarino:2015jca,Fluder:2015eoa}.

This proposal seems to be natural if one can think of this theory as the IR fixed point of the world volume theory on D2-branes probing a flat transverse space, deformed by a Chern-Simons term. The Chern-Simons term is induced onto D2-brane by the Romans mass. The theory has a superpotential of the form $\mathcal{W}={\rm Tr}X[Y,Z]$, which respects the same global symmetry as the supergravity solution. An $\cN=3$ theory can be obtained by an RG flow from the $\cN=2$ Chern-Simons matter theory triggered by a relevant deformation $\frac{1}{2}\epsilon{\rm Tr}Z^2$. In the IR, the coefficient $\epsilon$ runs to $k/\pi$, and the dimension of $Z$ becomes 1. Since the kinetic term of $Z$ becomes irrelevant, one can then integrate out $Z$, yielding a superpotential of the form $\mathcal{W}=\frac{\pi}{2k}{\rm Tr}[X,Y]^2$ for the remaining two massless chiral multiplets. This is the scenario studied in \cite{Gaiotto:2007qi}. The resulting IR theory consists of a Chern-Simon gauge field and a hypermultiplet encompassing an ${\rm SU}(2)_{\rm R}\times {\rm SU}(2)_{\rm f}$ global symmetry. The cosmological constant of the ${\cN}=3$ solution is smaller than that of the ${\cN}=2$ solution, therefore it is possible that in the gravity side, the ${\cN}=3$ solution can be attained from the $\cN=2$ solution via a holographic RG flow. If this is true, it is likely that the holographic dual of our $\cN=3$ solution is the $\cN=3$ Chern-Simons matter theory mentioned before.

Some evidence of this proposal can be seen from matching the bulk states of the 4D supergravity theory with the first few chiral primary multiplets in the dual CFT. A complete match between the short supergravity multiplets and the short CFT chiral primary multiplets requires the knowledge of the full Kaluza-Klein spectrum, nonetheless, we can make the first attempt by comparing the spectrum of the 4D supergravity theory with the lowest-level chiral primaries. Recall that \cite{Gallerati:2014xra} in the $\cN=3$ $AdS_4$ vacuum, the ${\cN}=8$ supergravity multiplet decomposes into one short supergravity multiplet ${\rm DS}(2,3/2,0)_{\rm S}$, two short gravitino multiplets ${\rm DS}(3/2,3/2,1/2)_{\rm S}$, one long gravitino multiplet ${\rm DS}(3/2,\sqrt{3},0)_{\rm L}$ and three massless vector multiplets ${\rm DS}(1,1)$. On the CFT side, the chiral primary multiplets (labeled by the quantum number of the chiral primary operator) of the previously mentioned $\cN=3$ CFT was studied in \cite{Minwalla:2011ma}. In their notation, the short supergravity multiplet corresponds to the short chiral primary multiplet $(\frac{3}{2},\frac{1}{2},0,0)$, the two short gravitino multiplets correspond to $(\frac{3}{2},0,\frac{1}{2},\frac{1}{2})$, and the three massless vector multiplets correspond to $(1,0,1,1)$.

As possible future research directions, it should be interesting to look for a D2 brane background whose near horizon limit smoothly approaches the $\cN=3$ solution found in the paper. Remember that besides the supersymmetric solutions, we have also found two stable nonsupersymmetric critical points which are invariant under ${\rm SO}(3)_R\times {\rm SO}(3)_D$ and ${\rm G}_2$ respectively. The stability of these solutions suggests their CFT duals may exist. It is therefore very intriguing to ask what kind of CFT could possibly be dual to these nonsupersymmetric $AdS$ backgrounds.

\begin{figure}[h]
\centering\includegraphics[width=8cm]{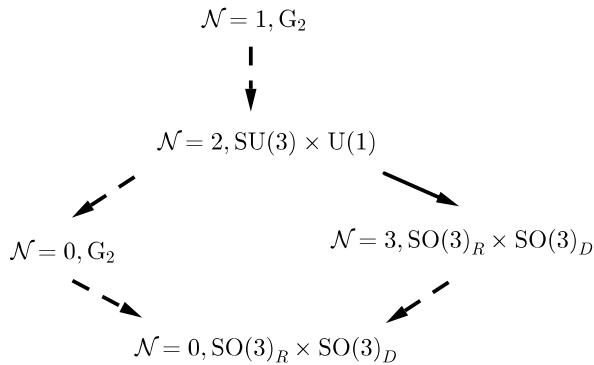}
\caption{List of critical points according to the heights of their cosmological constants. The $\cN=2$ point found in \cite{Guarino:2015jca} is included here. This suggests possible holographic RG flows among these critical points.}\label{rgflow}
\end{figure}

Diagram \ref{rgflow} exhibits the heights of the cosmological constants for various critical points, from which one can guess whether two critical points can be connected by a holographic RG flow. It should be interesting to see if one can actually construct domain wall solutions interpolating these critical points, which can be useful for studying the 3D CFT.
\section*{Acknowledgements}

  We are grateful to Chris Pope for useful comments.
This work is partially supported by DOE grant DE-FG02-13ER42020.

\end{document}